\documentclass{kluwer}
\usepackage{epsfig}
\usepackage{epsf}
\def\ltsima{$\; \buildrel < \over \sim \;$}
\def\simlt{\lower.5ex\hbox{\ltsima}}\begin{document}
\begin{article}
\begin{opening}
\title{The Accretion-Ejection Instability \\ and a ``Magnetic Flood'' 
scenario}            
\author{M.
\surname{Tagger}} \institute{Service d'Astrophysique (CNRS URA 2052), \\
CEA Saclay, 91191 Gif sur Yvette France}
\begin{abstract} 
I present a global view of recent results on the Accretion-Ejection
Instability (AEI), described in more details in other contributions to
this workshop.  These results address essentially the characteristics of
the AEI as a good candidate to explain the low-frequency QPO of X-ray
binaries, in particular (at $\sim 1-10$ Hz) of micro-quasars.  I then
discuss how, if the AEI is considered as the source of the QPO, a
possible scenario can be considered where the $\sim 30$ mn.  cycles of
GRS 1915+105 are controlled by the evolution of magnetic flux in the
disk.
\end{abstract}

\end{opening}

\section{Introduction}

The Accretion-Ejection Instability was first described by Tagger and
Pellat (1999), based on previous results of Tagger {\em et al.} (1991). 
It appears, in the inner part of disks threaded by a vertical (poloidal)
magnetic field, essentially as a spiral wave, similar to the ones of
galactic disks but driven by magnetic stresses rather than self-gravity. 
It becomes unstable by extracting energy and angular momentum from the
disk, causing accretion, and storing them in a Rossby vortex it
generates at its corotation radius (a few times the inner disk radius). 
The instability occurs if the field has a moderate amplitude (of the
order of equipartition), and presents a sufficient radial gradient.

We proposed the AEI as a possible explanation for the low-frequency
(``ubiquitous'') QPO observed in many X-ray binaries, and in particular
in micro-quasars.  This is based on the frequency of the instability (a
fraction of the rotation frequency at the inner disk radius, as
observed), and other properties such as its effect on the disk corona or
the fact that, as in galaxies, we expect it to form long-lived,
quasi-stationary spiral features, leading to a quasi-periodic modulation
of the disk emission.  I will present here a global view of recent work,
presented in more details in other contributions to these proceedings. 
I will then turn to the more prospective discussion of a ``magnetic
flood'' scenario for the $\sim$ 30 mn cycles of GRS 1915+105.  Starting
from the hypothesis that the QPO does correspond to the AEI, we consider
various observed properties of the QPO and of the disk, and confront
them with what we expect from the AEI. This leads us to this scenario,
compatible with known facts, although it cannot be proven at this stage;
considered as a possibility, and certainly not as the final word on
these cycles, it can thus be used as guidelines for our future work.

\section{New Results}
\subsection{Numerical simulations}
We have performed 2D ($r, \phi$) MHD simulations of a disk threaded by a
vertical (poloidal) magnetic field (Caunt and Tagger, these
proceedings).  The simulations globally confirm the linear analysis. 
The instability shows all the expected features (frequency, radial
propagation, vortex at its corotation radius, etc.).  Non-linearly it
saturates at high amplitude, giving a long-lived, quasi-stationary
spiral pattern in the disk.  Accretion causes the magnetic flux to pile
up in the central region, so that most simulations end up with an $m=1$
(single-armed) spiral, as expected.
\subsection{Emission of Alfv\'en Waves}
We have computed the flux of energy and angular momentum ``leaking''
from the vortex to the corona as Alfv\'en waves (Varni\`ere and Tagger,
these proceedings).  A future step will be to determine the way these
waves can be damped in the corona, ultimately powering a wind or a jet
from the accretion energy.
\subsection{QPO Frequency}
We expected the frequency of the $m=1$ mode to be quite sensitive to the
rotation profile in the disk.  In particular, relativistic effects when
the inner radius is close to the Last Stable Orbit allow the presence of
an Inner Lindblad Resonance (familiar in galactic dynamics).  We have
studied this in a pseudo-newtonian potential (Varni\`ere and Tagger,
these proceedings).  We find that, when the inner radius $r_{int} \simlt
1.4 r_{LSO}$, an ILR appears and reverses the correlation between
$r_{int}$ and the QPO frequency $\nu_{QPO}$.
\subsection{Comparison with observations}
We have studied (Rodriguez {\em et al.}, these proceedings) the
possibility that this could explain the reverse correlation between
$\nu_{QPO}$ and the color radius $r_{col}$, noted by Sobczak {\em et
al.} (2000) in GRO J1655-40 compared to other sources.  Comparing with GRS
1915+105, we find that this is a definite possibility, although present
results cannot prove it. However a detailed analysis of the
data leads to interesting results: \\
1) In the 30 mn cycles of GRS 1915+105, the QPO appears {\em before} the dip
in the X-ray curve: the QPO might be the cause, and not the consequence,
of the transition to the low state.  2) Some data from GRS 1915+105 are
compatible with the model of Merloni {\em et al.}, where anomalously low
measures of the color radius result from geometric effects in the disk
and corona; but this is not the case in GRO J1655-40, where low radii
correspond to high temperatures.  It rather indicates that the emitting
region is small and hot, {\em i.e.} that dissipation occurs in a spiral
shock or hot point in the disk, which could be a consequence of the AEI.
\section{Magnetic flood}
These results confirm that the AEI is a good candidate to explain the
``ubiquitous'' QPO. In order to guide our investigations, we have
attempted to go further by considering what this could tell us about the
disk.  This leads to a ``Magnetic Flood'' scenario (Tagger, 1999) where
the 30 mn cycles of GRS 1915+105 are controlled by the evolution of magnetic
flux in the disk.

We start from the fact that a magnetized disk is subject to the
Magneto-Rotational Instability (MRI), creating small-scale turbulence,
when the field is weak ($\beta=8\pi p/B^2> 1$), and to the AEI when it
is of the order of equipartition ($\beta\sim 1$).  The transition from
the high to the low state could thus correspond to a transition between
a weakly and a more strongly magnetized configuration, as magnetic flux
advected with the gas accumulates in the central region.  In a weak
field the MRI heats the disk by dissipation of the accretion energy;
when the field approaches equipartition the MRI stops and the AEI sets
in: this marks the appearance of the QPO, at the end of the high state. 
The heating of the disk decreases, because accretion energy is
redirected to the corona, so that pressure drops and makes $\beta$ even
lower: this would explain why the transition from the high to the low
state is so sudden (we note that in recent numerical simulations of the
MRI Hawley and Krolik (2000) find that, after some time, the magnetic
configuration becomes similar to the one we use here, and most of the
accretion energy is carried away by waves.  Technically these are fast
magnetosonic waves, {\em i.e.} the same spiral waves involved in the
AEI. Although the full 3D treatment limits the spatial resolution of
these simulations, we believe that this might very well be a first
manifestation of the magnetic flood).

Because of its decreased pressure the disk moves out, and then back in
as accretion from outer regions is still going on (the MRI is then seen
as the band-limited noise at lower frequencies, corresponding to larger
radii; this would explain the correlation between the break and QPO
frequencies).  The end of the process is marked by a reconnection event
(though other large-scale magnetic events could be considered) at the
inner edge of the disk, leading to the ejection of a plasmoid and to the
destruction of magnetic flux, allowing a return to the high state.

During the low state, some or most of the accretion energy from the
inner region of the disk is redirected to the corona.  Dissipation of
this energy, by a process we still have to work out, would then heat the
electrons, leading to the dominant comptonized tail seen in this low and
hard state.  This would also provide the link between disk accretion and
MHD jet models.

This leaves the question of what exists in the region between the disk
and the black hole.  An ADAF would be possible, but we also consider
another candidate: in the Blandford-Zjanek model, the black hole
contains magnetic flux due to external currents, held in a force-free
magnetic configuration between the black hole and the disk.  Such a
configuration, containing the magnetic flux advected to the central
``hole'' in our scenario, would be very similar to an ADAF in many ways:
it would form a fat torus of very low density plasma, where the heat
exchange time between ions and electrons is much longer than the
dynamical time.  The inner radius of the disk would be fixed by the
equilibrium with the magnetic pressure in the torus.

Although this scenario is still very schematic, it is compatible with 
the main observed properties of the 30 mn cycles of GRS 1915+105. It leads 
to a number of questions and prospects: the formation of the compact 
jet and of the ultra-luminous ejections, the plasma processes in the 
corona, and the longer term history of this source (which could be 
compared to other ones) are the most intriguing ones.

\end{article}
\end{document}